# Localized tip enhanced Raman spectroscopic study of impurity incorporated single GaN nanowire in the sub-diffraction limit

Avinash Patsha,[*] Sandip Dhara[*] and A. K. Tyagi

Surface and Nanoscience Division, Indira Gandhi Centre for Atomic Research, Kalpakkam,

India-603102

*Abstract*

The localized effect of impurities in single GaN nanowires in the sub-diffraction limit is reported using the study of lattice vibrational modes in the evanescent field of Au nanoparticle assisted tip enhanced Raman spectroscopy (TERS). GaN nanowires with the O impurity and the Mg dopants were grown by the chemical vapor deposition technique in the catalyst assisted vapor-liquid-solid process. Symmetry allowed Raman modes of wurtzite GaN are observed for undoped and doped nanowires. Unusually very strong intensity of the non-zone center zone boundary mode is observed for the TERS studies of both the undoped and the Mg doped GaN single nanowires. Surface optical mode of $A_1$ symmetry is also observed for both the undoped and the Mg doped GaN samples. A strong coupling of longitudinal optical (LO) phonons with free electrons, however is reported only in the O rich single nanowires with the asymmetric $A_1$(LO) mode. Study of the local vibration mode shows the presence of Mg as dopant in the single GaN nanowires.

Corresponding Authors Email : avinash.phy@gmail.com; dhara@igcar.gov.in

Apart from its advantage in detecting as well as imaging a single molecule,[1] noble metal (Au, Ag) coated tip enhanced Raman spectroscopy (TERS) is potentially used for localized study of vibrational properties of nanostructures in the sub-diffraction limit.[2,3] Raman scattering cross section is reported to be increased by $10^8$ times to facilitate single molecule detection in the evanescent field (plasmonics) of noble metal coated scanning probe microscopy tips.[4] The noble metal nanostructure is also responsible for focusing the excitation source in the sub-diffraction limit to probe the nanostructures in that dimension.[5,6] There are several remarkable reports of structural studies of single nanostructures of III-nitride using polarized micro-Raman spectroscopy,[7] and corresponding imaging techniques.[8,9] Without the influence of noble metal nanoparticle assisted plasmonics, a sub-diffraction limit is claimed for the crystallographic structural study of a single AlGaN nanowire using optical confinement of polarized light due to the dielectric contrast of nanowire with respect to that of surrounding media assisted with strong electron-phonon coupling of resonance Raman spectroscopy.[10] Leaking of polarization, however is a tricky issue in analyzing these results while working in the sub-diffraction limit without the help of plasmonics.[9] Hence, TERS study is expected to play a major role in analyzing localized chemical analysis in a single nanostructure of GaN. The intensity distribution of $E_2$(high) mode across a single GaN nanowire of diameter 200 nm was reported using TERS,[11] indicating the possibility of probing the surface and bulk of the individual nanowires by studying the variation of the symmetry allowed phonon modes. Subsequently, variations in the chemical composition, charge distribution, strain and the presence of stacking faults leading to different polymorphs close to the surface near the multiple quantum wells of single nonpolar InGaN/GaN nanorods were optically investigated using TERS with a spatial resolution in the sub-diffraction limit.[12] Variation of the In content for the InGaN layer of a few percent was reported with a lateral resolution < 35 nm. However, there are not many reports on the localized study of single III-nitride nanostructures in the sub-diffraction limit using TERS. At the same time, it is important to know the effect of doping, recognized either as unintentional (O, C) or intentional (Mg, Si), in single nanostructures for implementing them in a nanodevice. Unlike the case in thin films, variable pathways of dopant incorporation in nanowires lead to the nonuniform distribution of dopants along the radial or axial direction of a single nanowire.[13] Apart from the doping process, the type of dopant and the resultant defects in single nanowires with various sizes heavily influence the growth and electrical and optical characteristics of single nanowire devices.[14-18] Thus, it is essential to study the localized effects by dopants in single nanowires.

In the present report, we study the localized effect of O impurity and Mg doped single GaN nanowires in the sub-diffraction limit using TERS. Doped GaN nanowires were grown by the atmospheric pressure chemical vapor deposition (APCVD) technique in the catalyst assisted vapor-liquid-solid (VLS) process. A strong electron-phonon coupling is elucidated in the single GaN nanowires with O inclusion. At the same time local vibration mode (LVM) due to the Mg dopant in the Mg doped single GaN nanowires is also demonstrated.

Undoped and doped GaN nanowires were synthesized using APCVD technique in the VLS process. The optimized Au nanoparticles of size ~30 nm on Si substrate were used for the NWs growth. Ga metal (99.999%, Alfa Aesar) as precursor, $NH_3$ (99.999%) as reactant gas, and a mixture of ultra high pure (UHP) Ar+$H_2$ (5N) as carrier gases were used in the growth process. The Si substrate with Au nanoparticles was kept upstream of a Ga droplet in a high pure alumina boat (99.999%) which was placed into a quartz tube. The quartz tube was evacuated to a base pressure of $10^{-3}$ mbar until it attained a temperature which was just below the growth temperature of 900 °C to avoid contamination till the actual phase formation starts. The temperature of the quartz tube was slowly raised to an optimized growth temperature of 900 °C with 15 °C min$^{-1}$ ramp rate. The nanowires were grown for 60 min growth time by purging 10 sccm of $NH_3$ reactant gas and 20 sccm of Ar carrier gas. Incorporation of O was carried out at a oxygen partial pressures of $1.73 \times 10^2$ mbar (~$2 \times 10^5$ ppm $O_2$) by varying the base pressure of the growth chamber (quartz tube) and gas purity of $NH_3$ and Ar. Detailed growth process was reported elsewhere.[17] Mg doped nonpolar GaN nanowires were synthesized using $Mg_3N_2$ (Alfa Aesar) as a source for incorporating Mg and was reported earlier.[18] Au nanoparticles of size ~55($\pm$10) nm on Si substrate were used to grow the Mg doped GaN nanowires. Mg dopants, for *p*-type conduction in GaN NWs, were activated by thermally annealing the as-grown samples in a separate quartz tube in the $N_2$ atmosphere at 750 °C for 30 min.

The nanowires were separated from the growth substrate and dispersed on crystalline Cu(100) substrate for spectroscopic studies on the single nanowire. TERS measurements were carried out using the scanning probe microscopy setup (Nanonics, MultiView 4000; Multiprobe imaging system) coupled with a laser Raman spectrometer (inVia, Renishaw) in the backscattering configuration. The tip is atomic force microscopic (AFM) bent glass probe attached with Au particle (diameter < 100 nm) and operated it under non-optical normal force feedback. The Raman signals are collected by exiting the samples with 514.5 nm laser and are dispersed with a1800 gr.mm$^{-1}$ grating and a thermoelectric cooled charged couple device (CCD) detector. The spectra were collected using a 50X objective with numerical aperture (N.A.) value of 0.42. The nanowires with diameter in the range of 40-80 nm are well below the sub-diffraction limit of ~600 nm ($\lambda$/2N.A.) using an excitation wavelength ($\lambda$) of 514.5 nm.

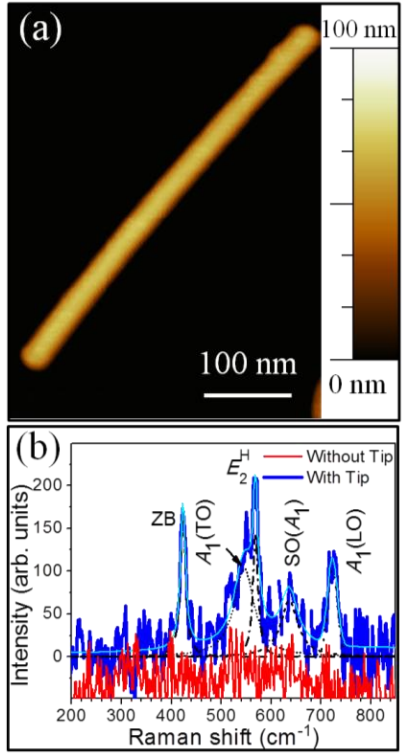

FIG. 1. (a) AFM image and (b) TERS spectra of undoped single nanowire without and with TERS tip.

The undoped nanowire, grown in the O free (<2 ppm)[17] condition, show smooth surface (Fig. 1(a)) with diameter ~ 40 nm in the absence of significant impurity. Typical TERS spectrum of undoped single nanowire with AFM tip (Fig. 1(b)) is deconvoluted using Lorentzian fitting, as the line width of the phonon spectra is a function of the phonon life time which follows the Lorentzian profile. The analysis shows the presence of the symmetry allowed modes at 537, 570 and 729 cm$^{-1}$ which correspond to the transverse optical $A_1$(TO), $E_2^H$ and longitudinal optical $A_1$(LO) phonon modes, respectively, of wurtzite GaN phase.[19] Crystallographic structural confirmation of the phase using high resolution transmission electron microscopic (HRTEM) analysis was reported in our earlier study.[17] A non-zone center zone boundary (ZB) phonon mode at 421 cm$^{-1}$ is also observed because of the finite size of the nanowire. Unusually very strong intensity of the ZB mode, which is comparable to the symmetry allowed mode is extremely rare and may be because of the fact that TERS study can be effectively used to invoke forbidden modes (due to plasmon interaction at the surface) using the evanescent field of noble metal nanoparticle.[20] Sharp $A_1$(LO) mode indicate absence of free carrier in the undoped nanowires. The $A_1$(LO) phonon peak characteristics are strongly influenced by the free carrier concentration (electrons/holes) in the GaN as a result of the LO phonon-free carrier coupling. Asymmetric broadening of the $A_1$(LO) phonon is observed as the free carrier concentration increases.[21,22] An additional strong mode at 633 cm$^{-1}$ is assigned as surface optical (SO) mode belonging to $A_1$ symmetry (SO($A_1$)) of GaN considering it in the two dimensional (2D) configuration.[23] The SO modes are generally observed when the translational symmetry of the surface potential is broken. Thus any perturbation of the surface potential induced by surface roughness, which is capable of absorbing the phonon momentum, is essential to make the SO phonon observable.[24] A localized probe in the evanescent field of Au nanoparticle with diameter <100 nm can easily focus light in the sub-diffraction limit of 20 nm.[25] Thus, approximation of the nanowire with diameter ~40 nm (sub-diffraction limit) in the 2D configuration with respect to the probe diameter of ~20 nm is reasonable. Observation of intense non-zone center SO($A_1$) phonon mode in a single GaN nanowire may be because of the presence of evanescent field in TERS measurement. In this context we may like to add that, apart from plasmonic activities of TERS and other metallic surface enhanced Raman spectroscopy (SERS), the similar observation of enhanced TO and the SO modes of 4-mercaptopyridine molecules on CdSe quantum dots, is also made for strong coupling of the charge-transfer transition and the exciton transition which are in resonance with the excitation energy.[26]

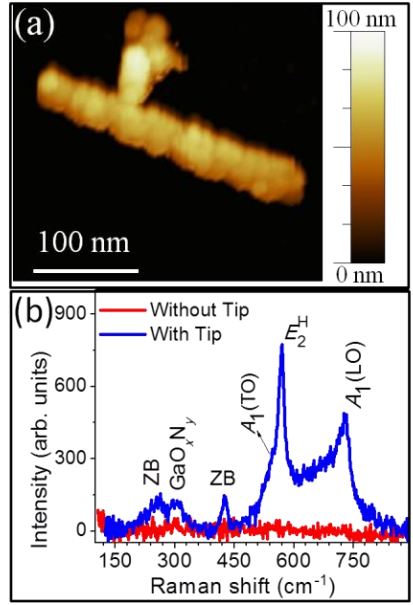

FIG. 2. (a) AFM image and (b) TERS spectra of O rich single nanowire without and with TERS tip.

The AFM image of a single nanowire with O inclusion (~2 ×10$^5$ ppm) shows non-uniform morphology (Figs. 2(a)). The nanowire diameter is in the range of 50 nm which is in the sub-diffraction limit. The corrugated morphology is because of inclusion of O in the lattice. As a matter of fact, highly inhomogeneous inclusion of Ga$_2$O$_3$ phase, using HRTEM analysis, were reported in our earlier study.[17] In order to observe the local chemical changes on single nanowires, TERS study of O rich sample was also recorded. The TERS spectrum with AFM tip shows symmetry allowed modes at 537, 569 and 729 cm$^{-1}$ corresponding to the $A_1$(TO), $E_2^H$ and $A_1$(LO) phonon modes, respectively, of wurtzite GaN phase.[19] Weakly intense ZB modes are also observed at 252 and 421 cm$^{-1}$. A tiny peak around 315 cm$^{-1}$ may be assigned to oxy-nitride (GaO$_x$N$_y$) phase.[23] Inhomogeneous inclusion of O may primarily be responsible for not identifying symmetry allowed modes of Ga$_2$O$_3$ which is found in our HRTEM study.[17] The absence of stoichiometric oxide phase in TERS study shows that O incorporation in GaN may be in the form of native defect occupying N site ($O_N$) or its complexes with N vacancy ($V_N$) contributing to generation of electron carrier in the nanowire.[21] As a matter of fact, the asymmetry in the $A_1(LO)$ mode is because of strong coupling between LO phonons and electron carrier generated by $O_N$ and $V_N$ in GaN.[21,22] Strong electron-phonon coupling and presence of native defects may influence also over the non-zone center phonons to make the ZB mode intensities weaker as compared to that observed for the nanowires with O deficient (<2 ppm) condition (Fig. 1(a)).

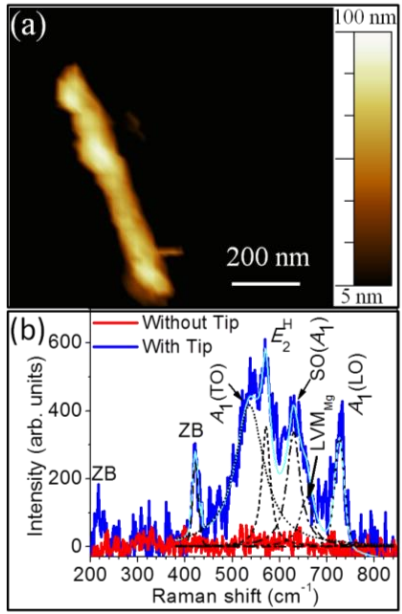

FIG. 3. (a) AFM image and (b) TERS spectra of Mg doped single nanowire without and with TERS tip.

The AFM image of single nanowires of Mg doped (GaN:Mg) sample is presented in Fig. 3(a) with fairly uniform morphology and diameter is in the sub-diffraction limit of around 80 nm. The Mg doping is reported to distort mainly the tip of the nanowire at the catalyst-nanowire interface where the dopants makes its way in the nanowire grown in the CVD technique.[18] In the TERS study with AFM tip for the GaN:Mg sample (Fig. 3(b)), additional peaks are observed at 633 and 656 cm$^{-1}$ along with ZB modes (252 and 421 cm$^{-1}$), and symmetry allowed $A_1$(TO) (537 cm$^{-1}$), $E_2^H$ (570 cm$^{-1}$) and $A_1$(LO) (729 cm$^{-1}$) modes of wurtzite GaN.[19] Sharp $A_1$(LO) mode also indicate absence of LO phonon-free carrier coupling, as discussed earlier in case of undoped sample. High intensities of ZB phonons, particularly for 421 cm$^{-1}$ mode, in the absence of free carrier-phonon coupling and any major native defects in the GaN:Mg sample support our argument for the observed weak intensities of the ZB modes in the O rich GaN nanowires (Fig. 2(b)). As discussed earlier in case of undoped sample (Fig. 1(b)), the peak at 633 cm$^{-1}$ corresponds to SO($A_1$) mode of GaN.[23] Stronger intensity of SO mode for the GaN:Mg sample than that for the undoped sample (Fig. 1(b)) may be because of the presence of dopant induced surface defect.[24] The additional peak at 656 cm$^{-1}$ corresponds to the local vibration mode (LVM) mode due to Mg doping in GaN.[21] An uniform and axial doping was claimed from detailed energy filtered TEM (EFTEM) imaging of GaN:Mg nanowires assisted with electron energy loss spectroscopy and x-ray photoelectron spectroscopy.[18]

In conclusion, tip enhanced Raman spectroscopy (TERS) study of the undoped and impurity (O and Mg dopant) incorporated single nanowires in the sub-diffraction limit show presence of wurtzite GaN phase. A strong longitudinal optical (LO) phonon-free carrier coupling is envisaged in O rich GaN nanowires from the observed asymmetry of the $A_1$(LO) mode where the generation of carriers are because of the presence of the N vacancy and O point defects. Surface optical modes originating from the surface defects are observed for both the undoped and Mg doped (GaN:Mg) samples. Unusually strong intensity of non-zone center zone boundary mode in the undoped and GaN:Mg nanowires show the role of evanescent filed in the absence of major native defects and free carrier-phonon coupling in these nanowires. Observation of local vibrational mode pertaining to Mg in GaN sample confirms the presence of Mg as dopant in the single GaN nanowires. Thus the TERS can be used for the localized effects by impurities in a single nanowire for developing the nanowire based electronic and optoelectronic devices.

One of us (AP) thanks DAE for allowing him to continue the research work. We thank S. Pal of Labindia Instruments Pvt. Ltd., India for valuable suggestions.